\documentclass[11pt]{article}
\usepackage[left=3cm, right=3cm, top=3cm]{geometry}
\pdfoutput=1

\begin{document}
\title{Droplets on a Tilted Plate}
\author{M. Musterd, V. van Steijn, C.R. Kleijn and M.T. Kreutzer \\
\\\vspace{6pt} Faculty of Applied Sciences, \\ Delft University of Technology, Delft, The Netherlands}
\maketitle
%% The abstract (in this file, and that submitted as text to arXiv) should
%include the exact phrase
%% "fluid dynamics video" or "fluid dynamics videos"
\begin{abstract}
In this short paper we present a fluid dynamics video of the deformation of droplets when tilted on a motorized stage. This is a submission to the 2013 Gallery of Fluid Motion which is part of the 66th annual meeting of APS-DFD. The video shows how differently placed droplets on the same surface will show a universal behaviour when tilted back and forth.
\end{abstract}
% main text
\section{Explanation and background of the video}
We deposit droplets in various initial states which are subsequently tilted to near roll-off and then tilted back to horizontal. We study the deformation from initial to final state. Droplets with different initial contact angles are achieved by depositing droplets of larger or smaller volume and emptying/filling these on the surface to the desired 50 $\mu$L. Emptying results in contact angles closer to the receding angle, whereas filling results in contact angles closer to the advancing angle. If necessary, vibration can be used to achieve the most stable contact angle on the given surface, which will be roughly half-way the advancing and receding angles.

When tilted the front or rear contact point of the droplet will move first, depending on the initial state of the droplet. When tilted close to the roll-off angle and then returned to horizontal the base length of all droplets, regardless of their initial shape, will have become similar. 

\section{Experimental setup and conditions}
All droplets are distilled water of MilliQ quality and have a volume of 50 $\mu$L. The surface consists of a 4-inch silicon wafer coated with a 10 micron layer of PDMS (Sylgard) in 5:1 ratio with the curing agent. This results in advancing and receding contact angles of 119 and 92 degrees respectively. The substrate is tilted at a rate of 0.25$^\circ/s$.

%\section{}

% The {\em hyperref} package is used to make links to the videos.
% %% The format is: \href{URL of video}{name that will appear in the text}
% Two sample videos are
% \href{http://ecommons.library.cornell.edu/bitstream/1813/8237/2/LIFTED_H2_EMS
% T_FUEL.mpg}{Video
% 1} and
% \href{http://ecommons.library.cornell.edu/bitstream/1813/8237/4/LIFTED_H2_IEM
% _FUEL.mpg}{Video
% 2}.
% It is recommended that the article include:
% \begin{enumerate}
% \item An explanation of what is shown in the video.
% \item The relevant conditions, parameters, etc..
% \item References to any papers containing further information on the
% videos.
% \item In the Abstract (in the LaTeX file and in the text submitted
% to arXiv), the exact phrase ``fluid dynamics video" or ``fluid
% dynamics videos". This is to facilitate subsequent searching.
% \end{enumerate}

% \section{Copyright Notice}
% This article has been published on the ArXiv under a perpetual, non-
% exclusive license. Copyright is retained by the authors. The attached video
% ﬁles are Copyright (c) 2013 by the authors and may not be copied, publicly
% presented or incorporated into any other derivative work without a clear
% attribution and written consent.

%
\end{document}